\documentstyle[preprint,tighten,epsf,aps,floats]{revtex}
\begin{document}
\hyphenation{col-li-der col-li-ders ca-lo-ri-me-ter ap-proxi-ma-tion
ca-lo-ri-me-ters re-nor-ma-li-za-tion mi-ni-mal di-rect ca-pa-bi-li-ty
ac-ce-le-ra-tor ac-ce-le-rates stan-dard exis-tence at-tri-bute at-tri-bu-ted
un-in-te-res-ting sta-tis-ti-cal in-fe-ren-ces ir-res-pec-tive dif-fe-rence
different hand-ling me-cha-nism me-cha-nics li-mits bran-ching le-vel scale
chan-nels spe-ci-fi-cal-ly dif-fer con-di-tio-nal maxi-mum li-near smal-ler
pro-ba-bi-li-ty bet-ween Higgs given scalar va-lues de-pends as-sump-tions
ge-ne-ra-tion in-clu-ding un-de-si-ra-ble cor-res-pon-ding boson bo-so-nic
la-bo-ra-to-ry ex-pe-ri-ment ana-ly-sis mea-sure-ment pro-ba-bi-li-ties
ex-pe-ri-men-tal ex-pe-ri-ence phy-sics dis-tri-bu-tion nor-ma-li-za-tion
po-la-ri-za-tion sta-tis-ti-cal-ly ca-te-go-ry ca-te-go-ries lep-ton leptons
de-pen-ding op-ti-mize op-ti-mized op-ti-mal in-ter-po-la-tion two small
effi-ci-en-cy phy-si-cal identifi-ca-tion results un-corre-la-ted spanned
luminosity se-ve-ral un-der-stand un-der-stan-ding uni-fied its 
ab-sent ag-rees all ana-lo-gous assu-ming beyond boun-daries boun-dary 
cal-cu-la-tions con-straint cor-ners current cha-rac-te-ris-tics 
de-cay de-pen-dence di-sap-pea-rance eva-lua-ted exa-mined ex-pects Fi-gure
ge-ne-ra-ting II ini-tia-ted iso-la-ted itera-tions jets li-mit 
means mo-de-ling near pa-ra-me-ter pa-ra-me-ters pri-ma-ri-ly re-cent Ref Refs
sca-lars si-mi-lar simu-la-tion SM space state Table using va-cu-um va-lue}
\def\mt{$m_t$}                          
\def\met{\mbox{${\hbox{$E$\kern-0.6em\lower-.1ex\hbox{/}}}_T$}} 
\def\etal{{\sl et al.}}                 
\def\isajet{{\sc isajet}}
\def\pythia{{\sc pythia}}
\newcommand{\ifm}[1]{\relax\ifmmode #1\else $#1$\hskip 0.15cm\fi}
\newcommand{\ie}{\mbox{i.e.}}
\newcommand{\lt}{\ifm{<}}
\newcommand{\gt}{\ifm{>}}
\newcommand{\tbar}{\ifm{\overline{t}}}
\newcommand{\ttb}{\ifm{t\tbar}}
\newcommand{\cstt}{\ifm{\sigma(\ttb)}}
\newcommand{\ltb}{\ifm{\log_{10}(\tan\beta)}}
\newcommand{\tb}{\ifm{\tan\beta}}
\newcommand{\mH}{\ifm{M_{H^{+}}}}
\newcommand{\tHb}{\ifm{t \rightarrow H^+b}}
\newcommand{\tWb}{\ifm{t \rightarrow W^+b}} 
\newcommand{\Htn}{\ifm{H^+ \rightarrow \tau^+ \nu}}
\newcommand{\Hcs}{\ifm{H^+ \rightarrow c \bar s}}
\newcommand{\HWbb}{\ifm{H^+ \rightarrow W^+ b \bar b}}
\def\ifm#1{\relax\ifmmode#1\else$#1$\fi}
\input{psfig}
\input{epsf}
\title {Search for Charged Higgs Bosons in Decays of Top Quark Pairs}
\author{                                                                      
B.~Abbott,$^{40}$                                                             
M.~Abolins,$^{37}$                                                            
V.~Abramov,$^{15}$                                                            
B.S.~Acharya,$^{8}$                                                           
I.~Adam,$^{39}$                                                               
D.L.~Adams,$^{49}$                                                            
M.~Adams,$^{24}$                                                              
S.~Ahn,$^{23}$                                                                
G.A.~Alves,$^{2}$                                                             
N.~Amos,$^{36}$
% ...
\\                                                                            
\vskip 0.70cm                                                                 
\centerline{(D\O\ Collaboration)}                                             
\vskip 1.0cm
}                                                                             
\address{                                                                     
\centerline{$^{1}$Universidad de Buenos Aires, Buenos Aires, Argentina}       
\centerline{$^{2}$LAFEX, Centro Brasileiro de Pesquisas F{\'\i}sicas,         
                  Rio de Janeiro, Brazil}                                     
\centerline{$^{3}$Universidade do Estado do Rio de Janeiro,                   
                  Rio de Janeiro, Brazil}                                     
\centerline{$^{4}$Universidad de los Andes, Bogot\'{a}, Colombia}             
\centerline{$^{5}$DAPNIA/Service de Physique des Particules, CEA, Saclay,     
                  France}                                                     
\centerline{$^{6}$Panjab University, Chandigarh, India}                       
\centerline{$^{7}$Delhi University, Delhi, India}                             
\centerline{$^{8}$Tata Institute of Fundamental Research, Mumbai, India}      
\centerline{$^{9}$Kyungsung University, Pusan, Korea}                         
\centerline{$^{10}$Seoul National University, Seoul, Korea}                   
\centerline{$^{11}$CINVESTAV, Mexico City, Mexico}                            
\centerline{$^{12}$Institute of Nuclear Physics, Krak\'ow, Poland}            
\centerline{$^{13}$Institute for Theoretical and Experimental Physics,        
                   Moscow, Russia}                                            
\centerline{$^{14}$Moscow State University, Moscow, Russia}                   
\centerline{$^{15}$Institute for High Energy Physics, Protvino, Russia}       
\centerline{$^{16}$University of Arizona, Tucson, Arizona 85721}              
\centerline{$^{17}$Lawrence Berkeley National Laboratory and University of    
                   California, Berkeley, California 94720}                    
\centerline{$^{18}$University of California, Davis, California 95616}         
\centerline{$^{19}$University of California, Irvine, California 92697}        
\centerline{$^{20}$University of California, Riverside, California 92521}     
\centerline{$^{21}$Florida State University, Tallahassee, Florida 32306}      
\centerline{$^{22}$University of Hawaii, Honolulu, Hawaii 96822}              
\centerline{$^{23}$Fermi National Accelerator Laboratory, Batavia,            
                   Illinois 60510}                                            
\centerline{$^{24}$University of Illinois at Chicago, Chicago,                
                   Illinois 60607}                                            
\centerline{$^{25}$Northern Illinois University, DeKalb, Illinois 60115}      
\centerline{$^{26}$Northwestern University, Evanston, Illinois 60208}         
\centerline{$^{27}$Indiana University, Bloomington, Indiana 47405}            
\centerline{$^{28}$University of Notre Dame, Notre Dame, Indiana 46556}       
\centerline{$^{29}$Purdue University, West Lafayette, Indiana 47907}          
\centerline{$^{30}$Iowa State University, Ames, Iowa 50011}                   
\centerline{$^{31}$University of Kansas, Lawrence, Kansas 66045}              
\centerline{$^{32}$Louisiana Tech University, Ruston, Louisiana 71272}        
\centerline{$^{33}$University of Maryland, College Park, Maryland 20742}      
\centerline{$^{34}$Boston University, Boston, Massachusetts 02215}            
\centerline{$^{35}$Northeastern University, Boston, Massachusetts 02115}      
\centerline{$^{36}$University of Michigan, Ann Arbor, Michigan 48109}         
\centerline{$^{37}$Michigan State University, East Lansing, Michigan 48824}   
\centerline{$^{38}$University of Nebraska, Lincoln, Nebraska 68588}           
\centerline{$^{39}$Columbia University, New York, New York 10027}             
\centerline{$^{40}$New York University, New York, New York 10003}             
\centerline{$^{41}$University of Rochester, Rochester, New York 14627}        
\centerline{$^{42}$State University of New York, Stony Brook,                 
                   New York 11794}                                            
\centerline{$^{43}$Brookhaven National Laboratory, Upton, New York 11973}     
\centerline{$^{44}$Langston University, Langston, Oklahoma 73050}             
\centerline{$^{45}$University of Oklahoma, Norman, Oklahoma 73019}            
\centerline{$^{46}$Brown University, Providence, Rhode Island 02912}          
\centerline{$^{47}$University of Texas, Arlington, Texas 76019}               
\centerline{$^{48}$Texas A\&M University, College Station, Texas 77843}       
\centerline{$^{49}$Rice University, Houston, Texas 77005}                     
}                                                                             
\maketitle
\begin{abstract}
We present a search for charged Higgs bosons in decays of pair-produced
top quarks using $109.2 \pm 5.8$ pb$^{-1}$ of data recorded from $p \bar p$ 
collisions at $\sqrt{s} = 1.8$~TeV by the D\O\ detector during 1992-96 at 
the Fermilab Tevatron.
No evidence is found for charged Higgs production, and most
parts of the [\mH,\tb] parameter space where the decay  \tHb\ has a branching
fraction close to or larger than that for \tWb\ are excluded at 95\% 
confidence level.
Assuming \mbox{\mt\ = 175 GeV} and 
$\sigma(p \bar p \rightarrow t \bar t) = 5.5$ pb, for \mH = 60~GeV, 
we exclude $\tb \lt 0.97$ and $\tb \gt 40.9$.
\end{abstract}
%
%
\newpage
The Higgs sector of the standard model (SM) consists of a single
complex doublet scalar field responsible for breaking electroweak
symmetry and generating gauge boson masses.
The simplest extension of the Higgs sector to two complex
doublets appears in many theories beyond the SM, including 
supersymmetry (SUSY).
Our study is based on the 
two-Higgs-doublet model where
one doublet couples to up-type quarks and neutrinos, and the other couples to
down-type quarks and charged leptons, as required by SUSY \cite{HHG}.
Under these circumstances, electroweak symmetry breaking leads to five
physical Higgs bosons: two neutral scalars $h^{0}$ and $H^{0}$, a neutral
pseudoscalar $A^{0}$, and a pair of charged scalars 
$H^\pm$.
The extended Higgs sector has two new parameters: \mH\ and \tb, where
\tb\ is defined as the ratio of the vacuum expectation values of the two 
Higgs fields.

Direct searches for $e^+e^- \rightarrow H^+H^-X$ at LEP have set lower
limits of 57.5--59.5~GeV on \mH\ at the 95\% confidence level (CL)
irrespective of \tb\ \cite{LEP}.
A measurement of the inclusive $b\rightarrow s \gamma$ decay rate gives
CLEO an indirect limit of $\mH \gt [(244+63/(\tb)^{1.3}]$ GeV, 
assuming only a two-Higgs-doublet extension to the SM~\cite{CLEO}.
From a measurement of the $b \rightarrow \tau \nu X$ branching fraction, 
ALEPH constrains $\tb / \mH < 0.52$ GeV$^{-1}$ at 90\% 
CL~\cite{ALEPH2}.
Based on a search for charged Higgs in decays of pair-produced top quarks
using hadronic decays of the $\tau$ lepton,
CDF has published limits in the [\mH,\tb] parameter space for $\tb \gt 5$ 
\cite{CDFch}.
Our search, also for $H^\pm$ in decays of \ttb, covers the entire range of
\tb\ in which leading order perturbative calculations are valid.

At leading order, the $H^+$ coupling to a down-type (up-type) quark
or neutral (charged) lepton is proportional to the fermion mass multiplied by 
\tb\ ($\cot\beta$).
The SM requires a $t$ quark to decay almost exclusively to a $W$ boson and a
$b$ quark, \ie, 
$B(\tWb) \approx 1$.
However, if $H^\pm$ exist with \mbox{$\mH < m_t - m_b$}, and
\tb\ is either very large or very small, then $B(\tHb)$ can be significant.
We assume $B(\tHb)+B(\tWb) = 1.$
For any given \tb, $B(\tHb)$ decreases as \mH\ increases.
It is further assumed that $M_{S^0}$  ($S^0 = h^0$, $H^0$, or $A^0$) are 
large enough for the decays $H^+ \rightarrow S^0W^+$ to be highly
suppressed for real or virtual $S^0$ and $W^+$ bosons.
Decays $H^+ \rightarrow V^0W^+$, where $V^0 = \gamma$ or $Z$, are absent at
the tree level~\cite{Rosiek}.
Hence, $H^+$ can only decay to fermion-antifermion pairs.
Consequently, if $\mH \lt m_t -m_b$, one might expect \Htn\ (favored if \tb\
is large) and \Hcs\ (favored if \tb\ is small) to be the only significant
possibilities.
Indeed, $B(\Htn) \approx 1$ if $\tb > 10$.
But if $\tb < 2$ and $\mH > 130$~GeV, then the large mass of the $t$
quark causes $B(H^+ \rightarrow t^\ast \bar b \rightarrow W^+b \bar b)$
to exceed $B(\Hcs)$~\cite{hwbb}.

Figure~\ref{fg:params} shows the region of the [\mH,\tb] plane examined in
this analysis.
The lower and upper boundaries on \tb\ (0.3, 150) are required for 
the applicability of perturbative calculations in $H^+$ Yukawa
coupling to $t$ and $b$ quarks.
The minimum for \mH\ is chosen at 50~GeV, somewhat below the most
recent lower limits from LEP.
This search is restricted to $\mH \lt 160$~GeV, somewhat less than $m_t - m_b$
(assuming $m_t = 175$~GeV); otherwise, the width of the charged Higgs
$\Gamma(H^+)$ becomes too large ($> 7.5$~GeV) near the upper boundary on
\tb, and leading-order calculations become unreliable.
For the same reason, $\Gamma(t)$ is required to be $<15$~GeV.
Since $\Gamma(\tWb) \approx 1.5$~GeV, irrespective of [\mH,\tb],
this amounts to requiring $B(\tHb) \le 0.9,$ and thereby excludes from our
analysis the dark-shaded regions at the two bottom corners of 
Fig.~\ref{fg:params}.
The cross-hatched regions correspond to $B(\tHb) \gt 0.5$.
Also shown in Fig.~\ref{fg:params} are the decay modes of $H^+$ that
dominate in different parts of the parameter space.
Analogous charge-conjugate expressions hold for $H^-$.
\begin{figure}
\centerline{\psfig{figure=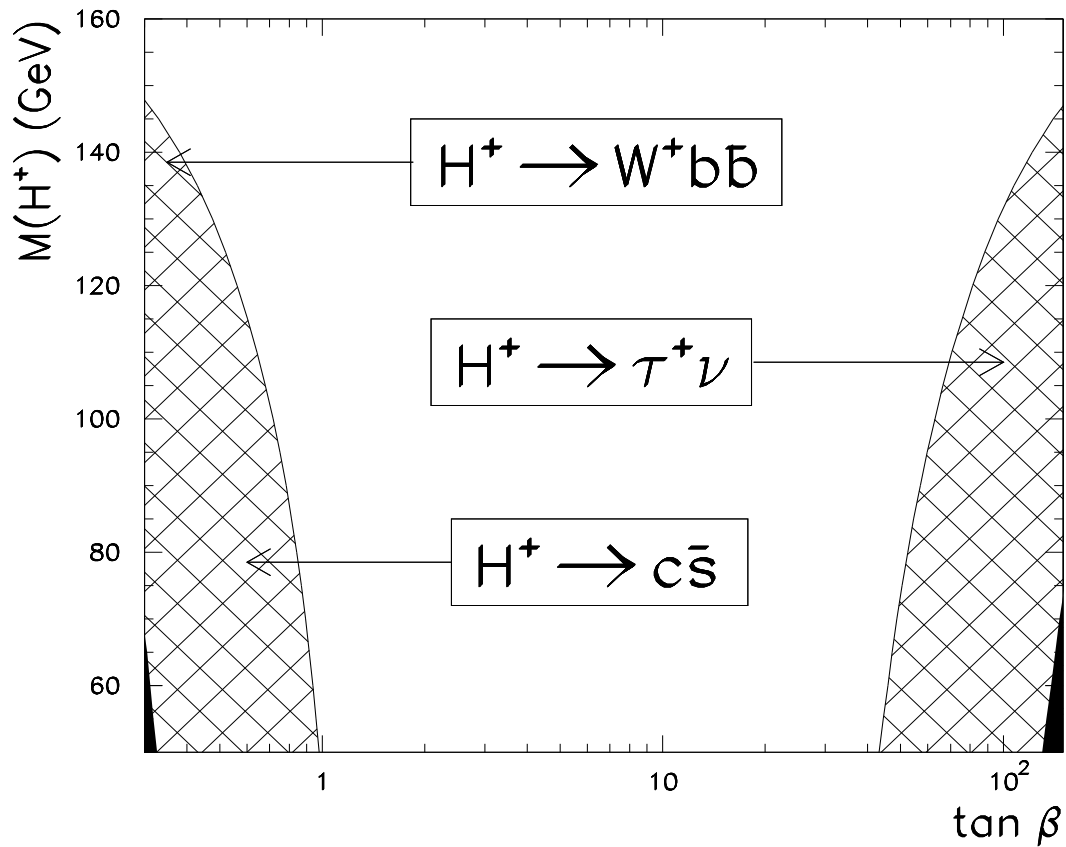,height=11.2cm,width=14.4cm}}
\vskip 2mm
\caption{The parameter space explored in this analysis.
Regions where $B(\tHb) \gt 0.5$ are shown cross-hatched, with the labels
for  various decay modes of the charged Higgs indicating their regions of 
dominance.
Regions where $B(\tHb) \gt 0.9$ (dark shaded areas) are not considered.}
\label{fg:params}
\end{figure}

For each top quark, there are four possible decay modes whose branching 
fractions depend on \mH\ and \tb: 
(1) \tWb; (2) \tHb, \Hcs; (3) \tHb, \HWbb; and (4) \tHb, \Htn.
If the decay mode of $t$ ($\bar t$) is denoted by $i$ ($j$), 
then the total acceptance for any set of selection criteria is given by
\begin{equation}
A(\mH,\tb) = {\displaystyle \sum_{i,j=1}^{4}} 
\epsilon_{i,j}(\mH)  B_i(\mH,\tb) B_j(\mH,\tb),
\label{eq:acc}
\end{equation}
where $\epsilon_{i,j}$ is the efficiency for channel $\{i,j\}$,
and $B_i B_j$ is the branching fraction.
All $B_i$ depend strongly on both \mH\ and \tb;
$\epsilon_{1,1}$ depends on neither, and all other $\epsilon_{i,j}$ depend on
\mH, but not on \tb.

\begin{table}
\caption{The $l$+jets and $l$+jets/$\mu$ event selection criteria.}
\begin{tabular}{c|cc}
 &
{$l$+jets} &
{$l$+jets/$\mu$} \\
\hline
$p_T(l)$ & \gt 20~GeV & \gt 20~GeV \\
$|\eta_{e(\mu)}|$ & \lt 2.0 (1.7)  & \lt 2.0 (1.7)  \\
\met & \gt 25~GeV & \gt 20~GeV \\
$E_T(j)$ & \gt 15~GeV & \gt 20~GeV \\
$|\eta_j|$ & \lt 2.0  & \lt 2.0  \\
\# of jets ($n_j$)& $\ge 4$ & $\ge 3$ \\
\# of $\mu$-tags & 0 & $\ge 1$  \\
Aplanarity & \gt 0.065 & \gt 0.040 \\
$H_T \equiv \sum_{i=1}^{n_j} E_T(j_i)$ & \gt 180~GeV& \gt 110~GeV\\
$p_T(l)+\met$ & \gt 60~GeV& -- \\
$|\eta(W)|$ & \lt 2.0 & -- \\
\end{tabular}
\label{tb:ljcuts}
\end{table}

A strong dependence of signal characteristics on the parameters of the model
makes 
an appearance search for signal difficult for us.
We therefore perform a disappearance search using selection
criteria optimized for the SM channel \{1,1\}.
One expects the efficiencies of these criteria for channels involving \tHb\
decays to be substantially different from that for channel \{1,1\}.
Consequently, if the assumption of $B_1 = 1$ leads to a measurement of 
the top quark pair production cross section \cstt\ in good agreement
with theoretical predictions, then those regions of the [\mH,\tb] parameter
space where $B_i$ is sufficiently large for any $i \ne 1$ can be excluded.
This strategy serves us well for $i=2$ and 4, but not for $i=3$.

The D\O\ detector is described in Ref.~\cite{d0detector}. 
We use the same reconstruction algorithms for jets, muons, and electrons as
used in our previous top quark analyses, and the same event 
selection criteria as for the measurement of \cstt\ in lepton+jets
final states~\cite{d0cs}.
These criteria are optimized for \ttb\ events where both top quarks decay
to $Wb$, with one $W$ decaying into $e \bar \nu$ or 
$\mu \bar \nu$, and the other into a $q \bar q^\prime$ pair.
The final state in such events is characterized by a high-$p_T$ isolated
lepton, large missing transverse energy (\met), and four jets.
The main sources of background are $W$+jets events and QCD multijet
events with a misidentified lepton and large \met.
Two of the jets in signal events are initiated by $b$ quarks.
A $b$ jet can be tagged by a muon contained
within the jet ($\epsilon B \approx 0.2$ per \ttb\ event).
Since such tagging is unlikely in background events, other requirements 
can be less restrictive for an event containing a $\mu$-tagged jet.
This class of $\mu$-tagged events is denoted by $l$+jets/$\mu$.
Events without a $\mu$-tagged jet, denoted by $l$+jets, are subject to
stricter requirements on kinematics.
Details of the selection criteria, summarized in Table~\ref{tb:ljcuts}, 
can be found in Ref.~\cite{d0cs}.
For $m_t = 175$~GeV, the selection efficiency for 
$\ttb \rightarrow W^+b W^- \bar b$ events is
3.42$\pm$0.11(stat)$\pm$0.55(syst) \%.
The jet energy scale, particle identification, and modeling of the signal
are the primary sources of systematic uncertainty.
The integrated luminosity, the number of observed events, and the expected
\ttb\ signal (assuming $B(\tWb) = 1$) and background 
are given in Table~\ref{tb:exptpars}.

\begin{table}
\caption{The integrated luminosity, the number of observed events, and the
expectations from background and SM \ttb\ signal (assuming \mt\ = 175~GeV,
\cstt\ = 5.5~pb), for $l$+jets and $l$+jets/$\mu$ selections combined.}
\begin{tabular}{c|c}
Integrated luminosity, $\cal L $ & $109.2 \pm 5.8$ pb\\
Estimated background, $n_B$ & $11.2 \pm 2.0$ \\
Expected signal (SM), $n_S$ & $19.7 \pm 3.5$ \\
Total events expected (SM) & $30.9 \pm 4.0$ \\
Events observed, $n_{\rm obs}$ & 30 \\
\end{tabular}
\label{tb:exptpars}
\end{table}

The measured values of \cstt\ \cite{d0cs,cdfcs}
and \mt\ \cite{mt} are based on the assumption of
$B(t \rightarrow W^+b) = 1$, and cannot be used in this analysis.
Hence, in our search, \cstt\ and $m_t$ serve as input parameters.
Production of \ttb\ takes place primarily via strong interactions, and
the cross section is not affected by the existence of $H^\pm$
(assuming no contribution from SUSY processes).
Calculations of \cstt\ based on 
QCD should therefore be reasonable \cite{Berger,Catani,Laenen}.
While there is no strong reason to use the measured value of $m_t$
when allowing $B(\tHb)$ to be large, lacking a compelling
argument in favor of an alternative choice, we use \mt~=~175~GeV.
A special version of \isajet~\cite{ISAJET} that includes the process
\HWbb\ is used for Monte Carlo simulation of \ttb\ events, and a similarly
modified version of \pythia~\cite{PYTHIA} for verification of the efficiencies.

Table~\ref{tb:exptpars} shows that the hypothesis of $B_1 \approx 1$ agrees
well with our experimental result.
Using Monte Carlo samples, the efficiencies and corresponding
uncertainties are calculated at several values of \mH,
and parametrized for each channel.
The efficiencies for all channels, for $\mH = 125$~GeV, are listed in
Table~\ref{tb:effs}.
The dependence of efficiency on \mH\ varies from channel to channel, but
efficiencies for a given channel rarely differ
by more than a factor of two over the range of \mH\ considered.
While $\epsilon_{2,2}$ is practically zero, $\epsilon_{1,3}$ and 
$\epsilon_{3,3}$ are close to $\epsilon_{1,1}$.
Consequently, we can exclude at a high level of confidence those regions
of parameter space where $B_2 \approx 1$ (small \tb, small \mH),
because, with almost no observable signal, it is extremely unlikely that an
expected background of $11.2 \pm 2.0$ events fluctuated to the
observed 30.
However, in regions where $B_3$ is comparable to or larger than
$B_1$ (small \tb, large \mH), the expected number of events is about the
same as that observed, and therefore such regions cannot be excluded. 
Low efficiencies for \ttb\ decays involving \Htn\ helps exclude
regions where $B_4$ is large (large \tb).

\begin{table}
\caption{The efficiencies $\epsilon_{i,j}$ of our selection criteria (in \%),
for \mt\ = 175~GeV and \mH\ = 125~GeV, for various decay modes of \ttb.
The row indices ($i$) denote: (1)~\tWb; (2)~\tHb,~\Hcs; (3)~\tHb,~\HWbb; and
(4)~\tHb,~\Htn.
The respective charge conjugate decays are denoted by the column indices 
($j$).}
\begin{tabular}{c|c c c c}
 & 1 & 2 & 3 & 4 \\
\hline 
1 & $3.42 \pm 0.56$ & $2.23 \pm 0.37$ & $3.35 \pm 0.61$ & $1.36 \pm 0.25$ \\ 
2 & $2.23 \pm 0.37$ & $0.04 \pm 0.01$ & $2.21 \pm 0.37$ & $1.07 \pm 0.20$ \\ 
3 & $3.35 \pm 0.61$ & $2.21 \pm 0.37$ & $3.71 \pm 0.67$ & $1.74 \pm 0.36$ \\ 
4 & $1.36 \pm 0.25$ & $1.07 \pm 0.20$ & $1.74 \pm 0.36$ & $0.41 \pm 0.09$ \\ 
\end{tabular}
\label{tb:effs}
\end{table}

For $n_{\rm obs}$ observed events, the joint posterior probability
density for \mH\ and \tb\ is given by
\begin{equation}
P(\mH,\tb|n_{\rm obs}) \propto  {\displaystyle \int  G({\cal L}) \int  G(n_B)
\int}  G(A) P(n_{\rm obs}|\mu) dA dn_B  d {\cal L},
\label{eq:n_obs}
\end{equation}
where $P(n_{\rm obs}| \mu)$, is the Poisson probability of observing
$n_{\rm obs}$ events, given a total (signal + background) expectation of
\begin{equation}
\mu(\mH,\tb)  =  A(\mH,\tb) \cstt {\cal L}  + n_B,
\end{equation}
and $G$ represents a Gaussian distribution.
The means and widths of the Gaussians for the integrated luminosity $\cal L$,
and the number of background events $n_B$, are given in 
Table~\ref{tb:exptpars},
while those for the acceptance $A(\mH,\tb)$, are calculated using
Eq.~(\ref{eq:acc}), with parametrized functions for 
$\epsilon_{i,j}$, 
and leading order calculations of  $B_i$, $B_j$.
\begin{figure}
\centerline{\psfig{figure=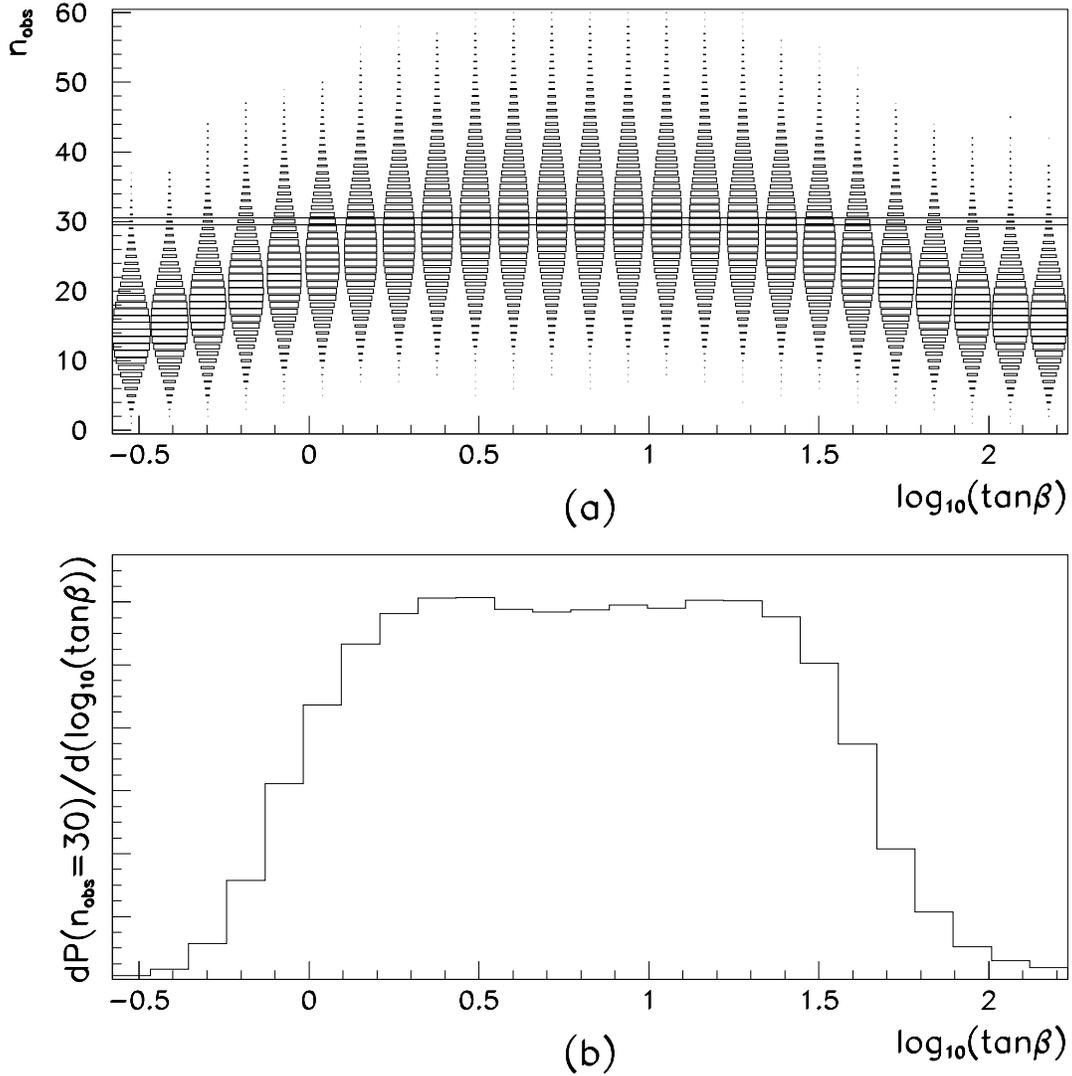,height=14.4cm,width=14.4cm}}
\caption{(a) Distribution of the number of  
Monte Carlo experiments in the $n_{\rm obs}$ {\it vs.} \ltb\ plane
for \mt\ = 175~GeV, $\cstt = 5.5$ pb, and \mH\ = 80~GeV.
(b) Posterior probabilty density for \tb, given the experimentally observed
value of $n_{\rm obs} = 30$ (the slice shown in (a)), for the above 
parameters.}
\label{fg:n_tb}
\end{figure}

Equation (\ref{eq:n_obs}), which we parametrize as a function of \mH\ and \tb,
gives a Bayesian posterior probability density for those 
parameters~\cite{Jaynes}.
The prior distribution is assumed to be uniform in \mH\ and in \ltb. 
Assuming instead that the prior is uniform in \mH\ and in $B(\Htn)$
does not significantly alter the posterior distribution.
To calculate probabilities, a Monte Carlo integration is carried out by
spanning the parameter space in steps of 5~GeV in \mH\ from 50~GeV to 160~GeV,
with 25 uniform steps in \ltb\ covering the range $0.3 < \tb < 150$ at each
value of \mH, and performing 200,000 trials of Eq.~(\ref{eq:n_obs}) at each
step.
The predicted probability for observing $n_{\rm obs}$ events, evaluated at
\mH\ = 80~GeV, for different values of \tb, is shown in Fig.~\ref{fg:n_tb}(a),
while Fig.~\ref{fg:n_tb}(b) shows the posterior probability density for \tb\
corresponding to $n_{\rm obs} = 30$, for \mH\ = 80~GeV. 
The 95\% CL exclusion boundary in the [\mH,\tb] plane is obtained by
integrating the probability density $P(\mH,\tb|n_{\rm obs})$,
given by Eq.~(\ref{eq:n_obs}), between contours of constant $P$.
The results, corresponding to $m_t = 175$~GeV, are shown in 
Fig.~\ref{fg:excl_cs} for three values of \cstt.
The largest value of \cstt\ (5.5 pb, with QCD resummation scale set to 
$m_t$~\cite{Berger}) yields the most conservative limits.
Tighter limits are set for smaller values of \cstt,
such as those given in Refs.~\cite{Catani,Laenen}.
Figure \ref{fg:excl_cs} also shows the result of a frequentist analysis of
our data wherein a point in the 
parameter space is excluded if
more than 95\% of the trials of Eq.~(\ref{eq:n_obs}) at that point 
yield $n_{\rm obs} < 30$.
Due caution must be exercised in comparing Bayesian and frequentist results
since the interpretation of ``confidence level'' is different between the two.
For a given value of \cstt, the excluded region increases with increasing
$m_t$ within the range 170 GeV $\lt m_t \lt$ 180 GeV, by an extent comparable
to that from a similar fractional decrease in \cstt\ at a fixed $m_t$.

To summarize, in a search for a charged Higgs boson that considers all of its
fermionic decay modes, we find no evidence of signal in the region of
$\mH \lt 160$ GeV, improve previous limits in the region of large \tb, and 
exclude a significant part of the previously unexplored region of small \tb.
Assuming $m_t = 175$ GeV and $\cstt = 5.5$ pb, $\tb < 0.97$ and $\tb > 40.9$
are excluded at 95\% CL for \mH = 60~GeV.
The limits become less stringent with increasing \mH.
Within the range of $0.3 \lt \tb \lt 150$, no lower limit
can be set on \tb\ for $\mH > 124$ GeV, and no upper limit for
$\mH > 153$ GeV.
A comparison between Figs.~\ref{fg:params} and \ref{fg:excl_cs} shows
that all regions of the [\mH,\tb] parameter space where $B(\tHb) \gt 0.45$,
except where $B(\HWbb)$ is large, are excluded at 95\% CL.

\begin{figure}
\centerline{\psfig{figure=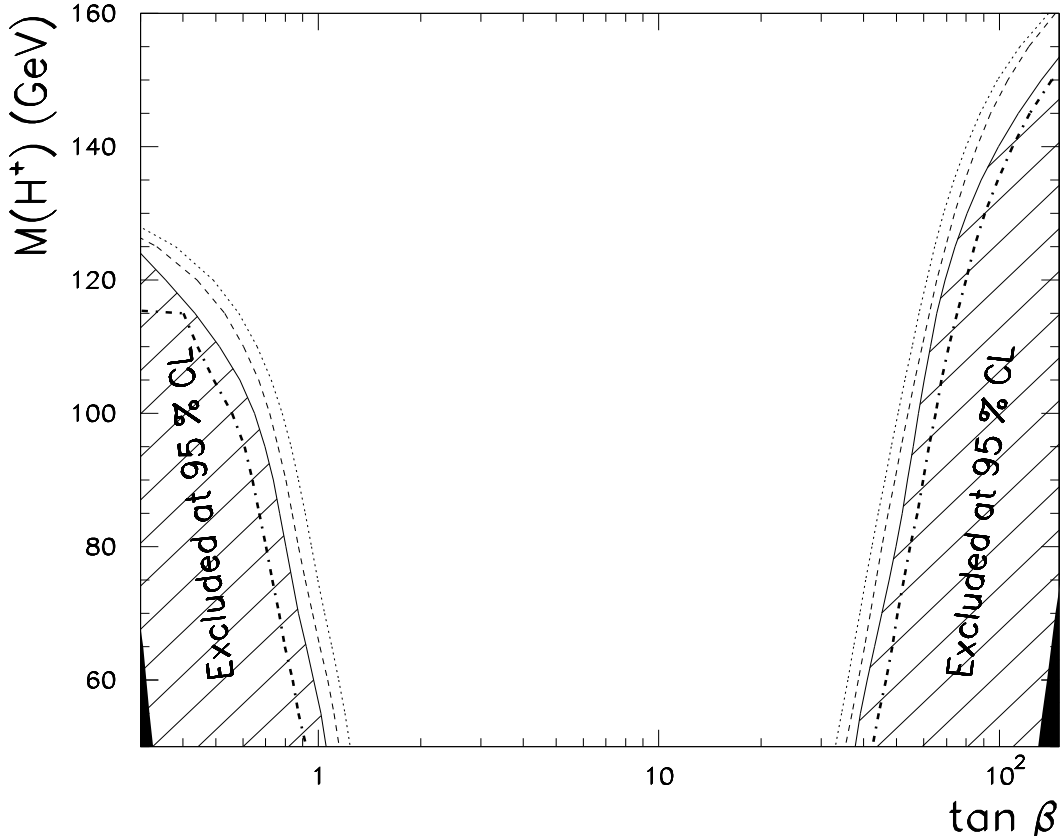,height=11.2cm,width=14.4cm}}
\caption{The 95\% CL exclusion boundaries in the [\mH,\tb] plane 
for  \mt\ = 175~GeV, and value of \cstt\ set to 5.5 pb (hatched area, 
solid lines), 5.0 pb (dashed lines), and 4.5 pb (dotted lines).
The thicker dot-dashed lines inside the hatched area represent the
exclusion boundaries obtained from a frequentist analysis with 
$\cstt = 5.5$ pb.}
\label{fg:excl_cs}
\end{figure}

We are grateful to D.~P.~Roy, J.~Wudka, and E.~E.~Boos for valuable
discussions on theoretical aspects of the analysis, and 
to S.~Mrenna for incorporating the process \HWbb\ into \pythia.
We thank the Fermilab and collaborating institution staffs for
contributions to this work and acknowledge support from the 
Department of Energy and National Science Foundation (USA),  
Commissariat  \` a L'Energie Atomique (France), 
Ministry for Science and Technology and Ministry for Atomic 
   Energy (Russia),
CAPES and CNPq (Brazil),
Departments of Atomic Energy and Science and Education (India),
Colciencias (Colombia),
CONACyT (Mexico),
Ministry of Education and KOSEF (Korea),
and CONICET and UBACyT (Argentina).


\begin{thebibliography}{999}

\bibitem[*]{ecuador}
Visitor from Universidad San Francisco de Quito, Quito, Ecuador.

\bibitem[\dag]{beijing}
Visitor from IHEP, Beijing, China.

\vskip 0.25cm

\bibitem{HHG}
J.~F.~Gunion, H.~E.~Haber, G.~Kane, and S.~Dawson,
``The Higgs Hunter's Guide'',
Addison-Wesley (1990).

\bibitem{LEP}
ALEPH Collaboration, CERN-EP/99-011 submitted to Phys. Lett. B;
L3 Collaboration, CERN-EP/98-149 submitted to Phys. Lett. B;
OPAL Collaboration, CERN-EP/98-173 submitted to Eur. Phys. J. C,
hep-ex/9811025.

\bibitem{CLEO}
CLEO Collaboration, M.~S.~Alam \etal, Phys. Rev. Lett.
{\bf 74}, 2885 (1995).

\bibitem{ALEPH2}
ALEPH Collaboration, D. Buskulic \etal, Phys. Lett. B {\bf 343}, 444 (1995).

\bibitem{CDFch}
CDF Collaboration, F.~Abe \etal, Phys. Rev. Lett.
{\bf 79}, 357 (1997).

\bibitem{Rosiek}
J.~Rosiek, Phys. Rev. D {\bf 41}, 3464 (1990).

\bibitem{hwbb}
E.~Ma, D.~P.~Roy, and J.~Wudka, Phys. Rev. Lett.
{\bf 80}, 1162 (1998).

\bibitem{d0detector} 
D\O\ Collaboration, S.~Abachi \etal, Nucl. Instrum. 
Methods A {\bf 338}, 185 (1994).

\bibitem{d0cs}
D\O\ Collaboration , B.~Abbott \etal, 
 Phys. Rev. Lett. {\bf 79}, 1203 (1997).

\bibitem{cdfcs}
CDF Collaboration, F.~Abe \etal, Phys. Rev. Lett. {\bf 80}, 2779 (1998).

\bibitem{mt}
D\O\ Collaboration, B.~Abbott \etal, 
 Phys. Rev. Lett. {\bf 79}, 1197 (1997);
CDF Collaboration, F.~Abe \etal, Phys. Rev. Lett. {\bf 80}, 2767 (1998).

\bibitem{Berger}
E.~L.~Berger and H.~Contopanagos, Phys. Rev. D {\bf 54}, 3085 (1996).

\bibitem{Catani}
S.~Catani, M.~L.~Mangano, P.~Nason, and L.~Trentadue,
Phys. Lett. B {\bf 378}, 329 (1996).

\bibitem{Laenen}
E.~Laenen, J.~Smith and W.~L. van Neerven,
Phys. Lett. B {\bf 321}, 254 (1994).

\bibitem{ISAJET}
F.~Paige and S.~Protopopescu, BNL Report BNL38034, 1986 (unpublished).
We used version 7.22.

\bibitem{PYTHIA}
T.~Sj\"{o}strand, Computer Physics Commun. {\bf 82}, 74 (1994).

\bibitem{Jaynes}
E.~T.~Jaynes, ``Probability Theory: The Logic of Science'',\\ 
{\it ftp://bayes.wustl.edu/pub/Jaynes/book.probability.theory/}
(unpublished).

\end{thebibliography}
\end{document}